# The Encoding of Natural Numbers as Nested Parentheses Strings with Associated Probability Distributions

**R. D. Ogden, ro01@txstate.edu**
Computer Science Department, Texas State University (retired)

***Abstract***

We provide an efficient encoding of the natural numbers {0,1,2,3,...} as strings of nested parentheses {(),(()),(()()),((())),...}, or considered inversely, an efficient enumeration of such strings. The technique is based on the recursive definition of the Catalan numbers. The probability distributions arising from this encoding are explored. Applications of this encoding to prefix-free data encoding and recursive function theory are briefly considered.

## 1. Introduction

In his version of algorithmic information theory [AT], Gregory Chaitin uses a version of the programming language Lisp to formulate, prove, and illustrate results similar to (but presumably stronger than) those of Gödel in undecidability theory. The language Lisp is based on lists represented by separated items enclosed by pairs of parentheses. In an effort to formulate an ultimate minimal version of Lisp, we developed a version based on parentheses alone; i.e., as certain strings on a two-letter alphabet such as $B$ ={'(' , ')' } ; we call this language BILL (i.e., BInary Little Lisp). To be capable of implementing any general recursive function and thus Turing equivalent, it was necessary to develop concrete ways of encoding names for infinitely many variables as well as for the natural numbers, at least. The result was the algorithm presented in this note and linked to a *Mathematica* program in an appendix. *Mathematica* was the principal investigative tool for this report.

We conclude this introduction by explaining certain terminology and notation. $\mathbb{N}$ is the set of natural numbers {0,1,2,3,...} considered as finite Von Neumann ordinals (e.g., 0 is the empty set $\phi$, 1={0}, 2={0,1}, 3={0,1,2}, etc.). The positive integers {1,2,3,...} = $\mathbb{Z}_+$. The cardinality of a set S is denoted by #S. A *list* is a function whose domain is a natural number; this includes the empty list $\Lambda$ whose domain is 0. The set of items on a list is the range of the function. If $A$ is a set and $n \in \mathbb{N}$, $A^n$ denotes the functions from $n$ to $A$. This set is also the set of lists of items from $A$ with $n$ elements; $A^n$ is also referred to as the strings of length $n$ on the alphabet $A$. . The set of all strings on the alphabet $A$ is denoted by $A^* = \bigcup_{n \in \mathbb{N}} A^n$ . If $x$ is a list (or a string) then $\#x$ is its length. $A^+$ denotes the set of non-empty strings on $A$.

If $\xi$ is a proposition then the expression $[\![ \xi ]\!]$ has value 1 if $\xi$ is true and 0 if $\xi$ is false.

# 2. The Language of Binary Symbolic Expressions

Binary symbolic expressions are essentially strings of nested parentheses; we shall abbreviate binary symbolic expression by bsx (plural: bsxes). As a language, the language may be described recursively in Backus-Naur form by

*bsxlist ::=* Λ | *<bsx><bsxlist> , bsx ::= (<bsxlist>)*

The bsx language is one of the simplest non-regular languages. Apparently, strings of type bsxlist form the simplest Dyck language studied by formal linguists. A more direct characterization is

*A bsx is a string x on the alphabet B*={ ( , ) } *such that for each j* ∈ *#x* , #{$i < j$ | $x_i$ = '('} > #{$i < j$ | $x_i$ = ')'}, *but* #{$i$ | $x_i$ = '('} = #{$i$ | $x_i$ = ')'}

In other words, at any point before the end of the string the number of left parentheses is greater than the number of right parentheses, but at the end the numbers must balance.

Thus a necessary (but certainly not sufficient) condition that a string on *B* be a bsx is that the string length be even, with an equal number of left and right parentheses. For our purposes the *size* of a bsx is a more convenient measure:

*The size of a bsx z = sz(z) = (#z - 2)/2*

The simplest bsx is" ()", which we denote by *nil* . It has stringlength 2 and size 0. Considered as a list on *B*, () is literally the set of ordered pairs {(0, '(' ), (1, ')' } but we rarely express bsxes in this way except to emphasize that they are lists of ordered pairs. When we are thinking of "()" as representing a list of lists, we refer to it as the empty list.

The Backus-Naur definition shows that bsxlists are built by prepending bsxes, starting with the empty list; then bsxes are made by enclosing bsxlists in parentheses. So besides being a mere list of parentheses, it represents a list of items, each of which is itself a bsx.

As an example, let *a* = (()(())((()())()())). The string length of *a* = 20, and the size of *a* = 9. *a* is built out of the list of three bsxes, namely (), (()), and ((()())()()). These bsxes are in turn built out of bsxlists, on down to (). The following is a representation of *a* as an ordered tree:

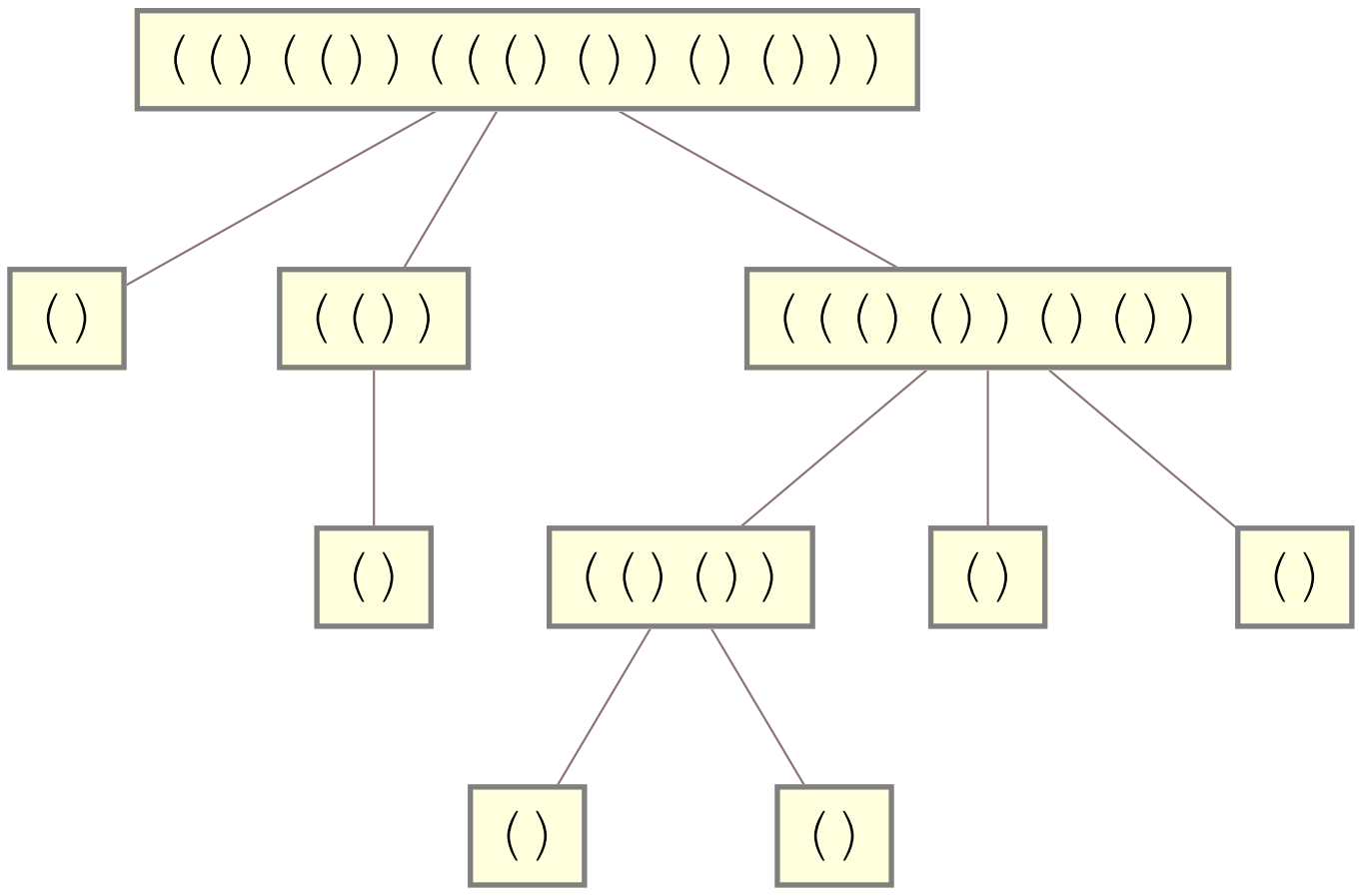

# 3. Join, the Head, and the Tail

If we consider a bsx $x = ( \xi_0\ \xi_1 \ldots \xi_{n-1} )$ as specifying a list of bsxes, then at least if the list is nonempty it makes sense to consider the first item $\xi_0$ . We call this first item the *head* of *x.* The rest of the list after the first item is removed, namely $( \xi_1 \ldots \xi_{n-1} )$, is called the *tail* of x. If $x = (\xi_0)$ has only one item then its tail is the empty list *nil*=() . We define both the head and tail of *nil* to be *nil* .For example, in the case *a* = (()(())((()())()())) we have *head*(*a*) = () and *tail*(*a*) = ((())((()())()())) .

Since our basic definition of bsxes was as strings, we should specify a separation into head and tail in terms of their identity as strings. The following simple algorithm shows how to mark the end of the head *h* as a substring of *x,* so letting the tail be the original string with the head excised.

```
Step 0. /* Algorithm to separate a bsx x into its head and tail */
      Input x; /*   x = (x[0], x[1], ..., x[# x - 1]) as a string */
    Step 1. If x == nil then return ((nil, nil)) and exit.
    Step 2. /* We know x starts out as  (( and we start n, the net ' (' count as 1
and i the position as 1 (x starts at 0 : -) */
            /* Initialize */ Let n = 1; Let i = 1
            Step 3. /* Move on and test*/  Let i = i + 1;
          If x[i] == "(" then Let n = n + 1 and repeat Step 3.
            Step 4. /* Since x is a bsx, x[i] must be a ')' */
         Let n = n - 1; If n > 0 go to Step 3.
       Step 5. /* The parentheses balance at the current position, i. */
            /*h is the head*/
            Let h = x[1 : i]; /* meaning the substring of characters from 1 to i */
        /*t is the tail*/
        Let t = x[0] <> x[i + 1 :] /* 1 st parenthesis followed by the substring
following position i to the end */
        Step 6. return ((h, t)) and exit.
```

The head and tail are fundamental for recursive algorithms involving bsxes. The inverse operation of joining two bsxes by prepending the first to the second. In terms of strings,

$$join(x,y) = y[0] <> x <> y[1:]\ .$$

The definition of *join* shows that as strings, #*join*(x,y) = #x + #y, but in terms of size,

$$sz(join(x, y)) = sz(x) + sz(y) + 1 \tag{1}$$

From this it follows that if we interpret bsxes as ordered trees as indicated above, then the size is one less than the number of nodes in the tree.

Joining and separation into head and tail are easily seen to be inverse operations, in that

$$head(join(x, y)) = x,\ tail(join(x, y)) = y,\ \text{ if } w \neq nil \text{ then } w = join(head(w), tail(w)) \tag{2}$$

# 4. Enumeration of Binary Symbolic Expressions

For $n \in \mathbb{N}$, let $\mathrm{BSX}_n$ denote the set of bsxes of size $n$ [equivalently, of string length $2n+2$], and let $\mathrm{BSX}_* = \biguplus_{n\in\mathbb{N}} \mathrm{BSX}_n$ be the set of all bsxes. Let $C_n = \#\mathrm{BSX}_n$ , the number of bsxes of size $n$. As it turns out, $C_n$ is the $n$th Catalan number. We derive this result in what follows, in a form suitable for our purposes. First, note that $\mathrm{BSX}_0 = \{\ ()\ \}$ and $\mathrm{BSX}_1 = \{\ (())\ \}$, so $C_0 = C_1 = 1$ .

Now suppose $n>1$, and $w \in \mathrm{BSX}_n$. By (2), $w = join\ (head(w), tail(w)) = join(x,y)$ for some $x$ and $y$. Since $n = sz(w) = sz(x) + sz(y) + 1$, clearly $w \in join[\biguplus_{k\in n} \mathrm{BSX}_{n-1-k} \lozenge \mathrm{BSX}_k]$ . So since the union is disjoint, $\#\mathrm{BSX}_n = \sum_{k\in n} \#\mathrm{BSX}_{n-1-k} \cdot \#\mathrm{BSX}_k$ , so if $n>0$, $C_n = \sum_{k\in n} C_{n-1-k}\, C_k$ . To cover all cases, we can write

$$C_n = [\![\, n == 0\, ]\!] + (C * C)_{n-1} \tag{3}$$

where $*$ is the convolution operation defined on sequences on $\mathbb{N}$ by $(a*b)_j = \sum_{k=0}^{j} a_{j-k} b_k$ .

The standard method for solving equations involving shifts and convolution is that of generating functions [see, e.g., [ConMath] ]. The generating function of the sequence $a$ is the formal power series defined by $G(a)(z) = \sum_{n\in\mathbb{N}} a_n z^n$ . The "operational calculus" of $G$ [a.k.a. the $Z$-transform] is fairly straightforward:

$$G(D(a))(z) = z\, G(a)(z),\ \text{where } D(a)_k = a_{k-1};\ \ G(\delta)(z) = 1,$$
$$\text{where } \delta_k = [\![k == 0]\!];\ \ G(a * b)(z) = G(a)(z)\, G(b)(z)$$

Now if we apply the $G$ operation to both sides of the equation $C = \delta + D(C * C)$, we get

$G(C)(z) = 1 + z \cdot (G(C)(z))^2$

Solving the quadratic equation for $G(C)(z)$, we get

$$G(C)(z) = \left(1 - \sqrt{1 - 4z}\right) / (2z) \tag{4}$$

When we expand $(1 - 4z)^{1/2}$ by the binomial theorem, after simplifying we get

$$G(C)(z) = 1 + 2 \sum_{n=1}^{\infty} \binom{1/2}{n+1} (-4z)^n$$

so for $n>0$, $C_n = (-1)^n 2^{2n+1} \binom{1/2}{n+1}$. So after applying standard binomial identities, we get the classic formula, valid for all $n \in \mathbb{N}$ :

$$C_n = \binom{2n}{n} \Big/ (n+1) \tag{5}$$

# 5. The Numeric Versions of Join, Head, and Tail

We now define numeric analogs of the string operations of head, tail, and join. But preceding that we must define the numeric analog of the size of a bsx.

*$S_n = \sum_{k=0}^{n-1} C_k = \sum_{k \in n} C_k$ is the sum of the first n Catalan numbers; $S_0=0$, $S_1=C_0=1$ .*

Next we define the analog of the logarithm for numbers, namely the size lgx:

$$\text{For } j \in \mathbb{N}, \text{ let } lgx(j) = \min\{n \in \mathbb{N} \mid j < S_{n+1}\}, \text{ so } 0 \le j - S_{lgx(j)} < C_{lgx(j)} \tag{6}$$

Thus each natural number $j$ has a pair of coordinates, $(n,r)$, related to $j$ by

$j = S_n + r$, where $0 \le r < C_n$ .

## The Numeric Separation into Head and Tail

Let $j > 0$ and let $n = \text{lgx}(j)$ . Then $n > 0$, and by (6), $0 \le x_0 < C_n$, where $x_0 = j - S_n$. Now from (3) we get $C_n = \sum_{k \in n} C_{n-1-k} C_k$ . Thus there is a unique value of $k$, $p$ say, such that $0 \le x_1 < C_q C_p$ where $x_1 = x_0 - \sum_{k \in p} C_{n-1-k} C_k$, and $p + q = n - 1$. Finally, let $h = S_p + \lfloor x_1 / C_q \rfloor$ , and $t = S_q + x_1 \bmod C_q$ . Clearly, $0 \le h - S_p < C_p$ and $0 \le t - S_q < C_q$, so $p = \text{lgx}(h)$ and $q = \text{lgx}(t)$.

$$j = S_{p+q+1} + \sum_{k \in p} C_{p+q-k} C_k + (h - S_p) C_q + (t - S_q) \tag{7}$$

We now set the numeric head of $j$, $nh(j) = h$ and likewise $nt(j) = t$. Formally,

*Define nh(0) = nt(0)=0 . For $j > 0$, let $n=$lgx$(j)$ and let $nh(j) = S_p + \lfloor (j - S_n - \sum_{k \in p} C_{p+q-k} C_k) / C_q \rfloor$, where p is uniquely determined by*

*$\sum_{k \in p} C_{n-1-k} C_k \le j - S_n < \sum_{k \in p+1} C_{n-1-k} C_k$, and $q = n - 1 - p$ .*

*Let $nt(j) = S_q + (j - S_n - \sum_{k \in p} C_{p+q-k} C_k) \bmod C_q$*

## The Numeric Joining

Let $a, b \in \mathbb{N}$; equation (7) suggests how to define the numeric join $nj(a,b)$.

*Let p=lgx(a) and q=lgx(b). Then nj(a,b) =* $S_{p+q+1} + \sum_{k\in p} C_{p+q-k}\, C_k + (a - S_p)\, C_q + (b - S_q)$

### Separation and Joining as Inverses

*Proposition (5.1): The map nj:* $\mathbb{N}\Diamond\mathbb{N} \to \mathbb{Z}_+$ *is one-to-one and onto, and its inverse is given by* $(nj)^{-1}(x) = (nh(x), nt(x))$*, for all x>0 .*

*Proof:* Let $g(x)=(nh(x),nt(x))$, for $x > 0$ . The proof will show that $g\circ nj$ = identity on $\mathbb{N}\Diamond\mathbb{N}$ and that $nj\circ g$= identity on $\mathbb{Z}_+$ .

Let $a,b \in\mathbb{N}$ , $p$ = lgx($a$), $q$=lgx($b$) , and let $x = nj(a,b)$ . We show $nh$(x) = $a$ and $nt(x) = b$ . Now

$0 \le (a - S_p)C_q + (b - S_q) \le (C_p-1)C_q + (C_q-1) = C_p\, C_q -1 < C_p\, C_q$, and so $x < S_{p+q+1}+ \sum_{k=0}^{p} C_{p+q-k}\, C_k \le S_{p+q+1}+C_{p+q+1} = S_{p+q+2}$, so $S_{p+q+1}\le x < S_{p+q+2}$, whence lgx($x$) = $p + q$ +1. Thus

$x - S_{p+q+1} - \sum_{k\in p} C_{p+q-k}C_k = (a - S_p)C_q+ (b - S_q)$, so the quotient upon division by $C_q$ is $(a - S_p)$ and the remainder is $(b - S_q)$. So according to the above definition, $nh(x) = a$ and $nt(x) = b$.

Next we show that $nj(nh(x),nt(x)) = x$, for all $x > 0$ . Well, according to the definition

$nj(nh(x), nt(x)) = S_{p+q+1} + \sum_{k\in p} C_{p+q-k}C_k + (nh(x) - S_p)C_q + (nt(x)-S_q)$, where $p$=lgx($nh(x)$) and $q$=lgx($nt(x)$). But then $p+q+1$ = lgx(x), and so by (7), $x = nj(nh(x),nt(x))$ .

This concludes the proof that numeric separation and numeric joining are mutual inverses.

# 6. The Encoding and Decoding of the Natural Numbers as Binary Symbolic Expressions

### Definitions of encoding *e* and decoding *d*

The encoding function $e: \mathbb{N} \overset{\text{iso}}{\to} BSX_*$ is defined recursively as follows:

$e(0)$ = "()", and if $x > 0$ then $e(x) = join(e(nh(x)), e(nt(x)))$

We prove $e$ is well-defined by induction on lgx($x$) .For $n$=0, the only number $x$ such that lgx($x$)=0 is $x$=0, and $e$(0) is defined to be "()" . So let $n > 0$ and suppose $e(t)$ is defined for all $s$ with lgx($s$) < $n$ . Suppose now lgx($x$) = $n$: from the definition of $nj$, lgx($nj(x)$) = lgx($nh(x)$) + lgx($nt(x)$) + 1, whence lgx($nh(x)$) < $n$ and lgx($nt(x)$) < $n$ . By the induction hypothesis, $e(nh(x))$ and $e(nt(x))$ are defined, and so the expression $join(e(nh(x)), e(nt(x)))$ for $e(x)$ is well-defined.

The decoding function $d$: $BSX_* \overset{\text{iso}}{\to} \mathbb{N}$ is defined recursively as follows:

$d$("()") = 0, and if $y \in BSX_*$ *and* $y \ne$ "()" then $d(y) = nj(d(head(y)), d(tail(y)))$

The proof that $d$ is well-defined uses induction on the size $sz(y)$ of $y$; it is very similar to the proof that $e$ is well-defined, applying equation (1) above.

### Proof that *d* and *e* are inverses

*Theorem (6.1):* $d = e^{-1}$

*Proof:* First, we show that $d(e(x)) = x$, by induction on $n$=lgx($x$) . So if $n$=0 we must have $x = 0$, so $d(e(0))=d$("()") = 0. Now let $n > 0$ and suppose the result is true for all $s$ with lgx($s$) < $n$. Let $x$ be such that lgx($x$) = $n$, so $x > 0$ . Then lgx($nh(x)$) and lgx($nh(x)$) are less than $n$ and so the induction hypothesis applies to them. Therefore

*d*(*e*(*x*)) = *d*(*join*(*e*(*nh*(*x*)), *e*(*nt*(*x*)))) [by above definition of *e*]

= *nj*(*d*(*e*(*nh*(*x*)), *d*(*e*(*nt*(*x*))))) [by above definition of *d*]

= *nj*(*nh*(*x*),*nt*(*x*)) [by induction hypothesis]

= *x* [by proposition (5.1)]

Now the proof that *e*(*d*(*y*)) = *y*, for all $y \in BSX_*$ is analogous, using induction on *sz*(*y*), the size of *y*. So, if *sz*(*y*) = 0 then *y* = "()", so *d*(*y*) = 0 and *e*(0) = "()", so the base case is established. So suppose $n > 0$ and *e*(*d*(*y*)) = *y* is true for all *y* with *sz*(*y*) < *n*. Now let $w \in BSX_*$ with *sz*(*w*) = *n*. Then

*e*(*d*(*w*)) = *join*(*e*(*nh*(*d*(*w*))),*e*(*nt*(*d*(*w*)))) by definition of *e*. But by definition, *d(w)* = *nj*(*d*(*head*(*w*)),*d*(*tail*(*w*))), so by proposition (5.1)

*nh*(*d*(*w*)) = *d*(*head*(*w*)) and *nt*(*d*(*w*)) = *d*(*tail*(*w*)) . Therefore

*e(d(w))* = *join*(*e*(*d*(*head*(*w*))), *e(d((tail*(*w*)))

= *join*(*head*(*w*), *tail*(*w*)) [by the induction hypothesis]

= *w* [by (2)]

Together these two results show that *d* and *e* are one-to-one and onto, and inverses of each other.

## ■ Consequences and Examples

The following brief table exhibits some examples of the encoding and suggests some conjectures.

| ***x*** | ***e(x)*** | lgx***(x)*** | ***#e(x)*** |
|---|---|---|---|
| 0 | () | 0 | 2 |
| 1 | (()) | 1 | 4 |
| 2 | (()()) | 2 | 6 |
| 3 | ((())) | 2 | 6 |
| 4 | (()()()) | 3 | 8 |
| 5 | (()(())) | 3 | 8 |
| 6 | ((())()) | 3 | 8 |
| 7 | ((()())) | 3 | 8 |
| 8 | (((()))) | 3 | 8 |
| 9 | (()()()()) | 4 | 10 |
| 10 | (()()(())) | 4 | 10 |
| 11 | (()(())()) | 4 | 10 |
| 12 | (()(()())) | 4 | 10 |
| 13 | (()((()))) | 4 | 10 |
| 14 | ((())()()) | 4 | 10 |
| 15 | ((())(())) | 4 | 10 |
| 16 | ((()())()) | 4 | 10 |
| 17 | (((()))()) | 4 | 10 |
| 18 | ((()()())) | 4 | 10 |
| 19 | ((()(()))) | 4 | 10 |
| 20 | (((())())) | 4 | 10 |
| 21 | (((()()))) | 4 | 10 |
| 22 | ((((())))) | 4 | 10 |
| 23 | (()()()()()) | 5 | 12 |

*Proposition (6.1): Let sz be the size function for bsxes as defined in Section (2), and let the numerical size function lgx be defined by (6) .*

*Then lgx(d(x)) = sz(x) for all x∈ BSX$_*$ and sz(e(y)) = lgx(y) for all y ∈ ℕ .*

*Proof:* Note that Theorem (1) implies that the equalities are equivalent, so let us establish *sz*(*e*(*y*)) = lgx(*y*) for all *y* ∈ ℕ by induction on lgx(*y*). The result is obviously true for 0; so suppose the result is true for all *w* ∈ ℕ with lgx(*w*) < *n*, and let lgx(*y*) = *n*. Then

*sz*(*e*(*y*)) = *sz*(*join*(*e*(*nh*(*y*)),*e*(*nt*(*y*))) [definition of *e*]

= *sz*(*e*(*nh*(*y*))) + *sz*(*e*(*nt*(*y*))) + 1 [by (1) in Section 2.]

= lgx*(nh(y))* + lgx*(nt(y))* + 1 [applying the induction hypothesis twice]

= lgx*(y)* [applying (7) above]

Next we consider the marginal cases, and prove the following propositions suggested by the above table:

*Proposition (6.2): Let $C_n$ be the nth Catalan number. Then for all x ∈ ℕ,*

*(6.2a) nj(0,x) = $C_{lgx(x)}$ + x, and if e(x) = (ξ ), where ξ is a bsxlist, then e(nj(0,x)) = (()ξ )*

*(6.3b) nj(x,0) = $C_{lgx(x)+1}$ + x, and if e(x) = (ξ ), where ξ is a bsxlist, then e(nj(x,0)) = ((ξ )) .*

Recall $S_n$ is the sum of the first *n* Catalan numbers, starting with $C_0$ = 1. First, apply the definition of *nj*, replacing *a* by 0 and *b* by *x*, and we get *nj(0,x)* = $S_{0+q+1} + \sum_{k\in 0} C_{q-k} C_k + (0-S_0) C_q + x - S_q = S_{q+1} + x - S_q = C_{lgx(x)} + x$, since *q* = lgx(*x*) . Next, by definition and theorem (6.1), *e*(*nj*(*0*,*x*)) = *join*(*e*(*0*),*e*(*x*)) = *join*( "()", (ξ)) = (()ξ) by definition of *join*. This proves (*a*) .

To prove (***b***), apply the definition of *nj*, replacing *a* by *x* and *b* by 0, and we get *nj(x,0)* = $S_{p+0+1} + \sum_{k\in p} C_{p-k} C_k + (x - S_p) C_0 + (0-S_0) = S_{p+1} + \sum_{k\in p} C_{p-k} C_k + x - S_p = C_0 C_p + \sum_{k\in p} C_{p-k} C_k + x = (C * C)_{p+1-1} + x = C_{\mathrm{lgx}(x)+1} + x$, using (3) and *p* = lgx(x) . Next, by definition and theorem (6.1), *e*(*nj*(*x*,*0*)) = *join*(*e*(*x*),*e*(*0*)) = *join*((ξ),"()") = ((ξ)) by definition of *join*. This proves (*b*) .

*Corollary (6.1) Let f(x) = nj(0,x) and $f^{\circ n}(x)$ = f applied to x n times. Then $f^{\circ n}(x) = S_{p+n} + (x - S_p)$, where p = lgx(x) . In particular, $f^{\circ n}(0) = S_n$ , and e($S_n$) = "(()...())" , which is n pairs "()" enclosed between '(' and ')' .*

*Proof:* The corollary follows easily by induction on *n*, once we observe that

$S_{p+n} \leq S_{p+n} + (x - S_p) < S_{p+n} + C_p \leq S_{p+n} + C_{p+n} = S_{p+n+1}$ , so that lgx($S_{p+n} + (x - S_p)$) = *p* + *n*, whence

$f(S_{p+n} + (x - S_p)) = C_{p+n} + S_{p+n} + (x - S_p) = S_{p+n+1} + (x - S_p)$ .

So, just as lgx is the analog of logarithms, so the sums of the Catalan numbers are the analogs of exponentials. The bsx "(()...())", where there are *n* pairs "()" is the first bsx of size *n* and string-length 2*n*+2 ; in the context of bsxes, that bsx is the *n*th power.

# 7. BSXes Considered for Prefix-free Serial Data Encoding

Suppose we have a channel of some sort connecting Alice and Bob, and the channel can be in only two states, 0 and 1. [In this section we will often use 0 for '(' and 1 for ')' .] The channel is normally in state 1. Now if Alice wants to send Bob a message, the message obviously has to be encoded somehow in binary, and then the bits have to be sent over the channel one by one. We assume that the time $\delta$ to send a bit is fixed whether the bit is 0 or 1, but $\delta$ is not necessarily known in advance; Alice and Bob have no common clock to synchronize the transmission. We also assume that nothing is known about the statistical character of the messages that Alice might send, nor at what point in time she might deign to send Bob a message

Several questions come up: How does Bob know the baud rate = $1/\delta$ ? when a message begins? or when it ends?

We can provide straightforward answers to the first two questions by establishing a simple protocol. Since the channel is normally in the 1 state, we will signal the start of a message when the channel drops to 0. That will start a timer of Bob's, which will stop when the channel returns to 1. The timer thus records $\delta$ for Bob. Next, the protocol requires that Alice send a 1 (with the same duration $\delta$, of course). So every message has a fixed two-bit overhead and a minimum transmission time of 2 $\delta$. Meanwhile, Bob waits for time $\delta/2$ and then samples the channel. If he receives another 0, he assumes he's just getting random noise, waits for the channel to return to 1, and then clears the timer and starts all over waiting for the channel to fall to a 0. But if he receives a 1, he assumes he's getting a message from Alice and starts sampling every $\delta$ time units.

But how does Bob know when the message ends? How can he tell a message which ends in a 1 from the channel returning to its normal state? Clearly, Alice and Bob must agree on a binary encoder/decoder, and the code must be such that you can tell when the codeword ends; i.e., the code must be prefix free. Let us formalize these considerations, assuming (to level the playing field) that the messages are natural numbers.

*A binary encoder/decoder pair B is a pair of maps* $(B_E, B_D)$ *with* $B_E: \mathbb{N} \xrightarrow{1-1} 2^+$ *and* $B_D: 2^+ \xrightarrow{onto} \mathbb{N}$ *such that* $B_D \circ B_E$ *= identity map on* $\mathbb{N}$.*. The set of codewords of B is the range of the encoder* $B_E$ *. The code B is prefix free provided there do not exist natural numbers* $a \neq b$ *such that* $B_E(b) = B_E(a) <> \xi$ *for some* $\xi \in 2^+$ *[recall <> denotes the concatenation of strings].*

*For the code B the function* $L_B(x) = \# B_E(x)$, the length of the codeword encoding $x \in \mathbb{N}$. *The code B is monotone if* $L_B$ *is a non-decreasing function.*

*Let* $\lambda_B = \lambda$ *be the enumeration of the range of* $L_B = \{\lambda_k \mid k \in \mathbb{N}\}$, $1 \leq \lambda_0 < \lambda_1 < \lambda_2 < \ldots$.

*Let* $\nu_B = \nu$ *be the sequence* $\nu_k = \# \{x \mid L_B(x) = \lambda_k\}$, the number of codewords of length $\lambda_k$ .

The following lemma is almost obvious:

*Lemma: If B is a monotone prefix – free code, then* $lg_2(x) \leq L_B(x)$

*Proof*: Since $L_B$ is monotone, $\{x \mid L_B(x) = \lambda_k\} = [A_k, A_{k+1}[ \; = \{x \in \mathbb{N} \mid A_k \leq x < A_{k+1}\}$ , where $A_k = \sum_{j \in k} \nu_j$ .

So given $x \in \mathbb{N}$ , let $L_B(x) = \lambda_k$, so $x + 1 \leq A_{k+1} = \sum_{j=0}^{k} \nu_j$ . It is clear that $\nu_k \leq 2^{\lambda_k}$, but in fact $\nu_k < 2^{\lambda_k}$ . For assume $\nu_k = 2^{\lambda_k}$ Then each string of length $\lambda_k$ is a codeword, and so there could be no codewords longer than $\lambda_k$ since its first $\lambda_k$ characters would always be a codeword, contradicting that B is assumed prefix-free. Therefore

$A_{k+1} = \sum_{j=0}^{k} \nu_j \leq \sum_{j=0}^{k} (2^{\lambda_j} - 1) \leq 2^{\lambda_k + 1} - 1 - (k+1) < 2^{\lambda_k + 1}$ , so

$\lg_2(x) \leq \log_2(x+1) \leq \log_2(A_{k+1}) < \lambda_k + 1$, whence $\lg_2(x) \leq \lambda_k = L_B(x)$ .

Thus the best we can hope for a monotone prefix-free code $B$ is that as $x \to \infty$ , $L_B(x) \sim \lg_2(x)$ . [We use the notation $a_n \sim b_n$ to mean $\lim_{n \to \infty} \frac{a_n}{b_n} = 1$ ].

One well-known example of such a code is the so-called Elias delta code http://en.wikipedia.org/wiki/Elias_delta_coding . For this code the codeword length is

$L_B(x) = 1 + 2 \lg_2(\lg_2(x)) + \lg_2(x)$ . Typically, codes of this type prepend a number of bits to specify the count of the $\lg_2(x)$ bits of the binary representation of $x$ , the number of added bits growing slower than $\lg_2(x)$. The BSX encoding achieves near optimal compression, but by a very different type of encoding.

## ■ BSX Encoding and the Rate of Growth of $S_n = \Sigma_{k<n} C_k$

The BSX encoder is the map $X_E : \mathbb{N} \xrightarrow{1-1} 2^+$ such that $X_E(a) = e(a)_1 <> \ldots <> e(a)_L$ ;that is, $X_E(a)$ is merely $e(a)$ with the leading 0 removed; this is to remove an unnecessary redundancy. Here $e$ is the encoder from $\mathbb{N}$ onto $BSX_*$ , and $L = 2\lg x(a) + 1$ is the codeword length.The codewords, the range of $e$, are precisely those bit strings which become bsxes when a "0" is prepended [ the bsxes considered as a subset of $2^*$, if we replace '(' by 0 and ')' by 1.]. The codewords for 0, 1, 2, 3 respectively are 1, 011, 01011, 00111 .

The design of a good decoder, using error correction based on a knowledge of the statistics of the source, would be a worthy project. One could base a decoder directly on the recursive definition of $d$, but such a decoder would require having the complete bitstring before decoding could begin. The following is an algorithm for an instantaneous decoding that works as the bits of the codeword are received serially. The practicality of this decoder depends, among other things, on whether the computations required by the loop in Step 3 can be carried out within time $\delta$, the bit duration.

---

```
Step 0. /* Algorithm to decode a bitstring as a bsx, where the input bits are
processed serially in “real time” as they are read. Considered as the code for an
ordered bare tree, the nodes are transversed in pre-order. */
      Input arguments:
       x; /*   x = (x[0], x[1], ..., x[# x - 1]) as a string of 0s and 1s */
      Output value:
       v; /* decoded returned value */
      Local variables:
       m, /* string length of x */
       p, /* index of current position in string */
       n, /* number of 0s (lps) thus far */
       b, /* bit currently input from string x */
       stk; /* stack of left parentheses 0s and computed values */
Step 1. Let m = #x; If m = 0 then return(0) and exit. Otherwise, Let p = 0 and Let
b = x[p];
        If b ≠ 0 then return(0) and exit. Otherwise,
        Let stk={-1,-1}; Let n = 2; Let v = 0.
Step 2. /* main loop to read bits */
        If p ≥ m  then Let v = 0; Go to Step 6. Otherwise, Let p =  p + 1; Let b =
x[p].
        If b = 0 then /* found another lp */
            Let n = n + 1; Push -1 onto stk and go back to Step 2.
Step 3. /* b is 1 */
        If stk is empty then Push v onto stk and go to Step 2. Otherwise,
        If Top[stk] ≥ 0 then Let v = nj[Top[stk],v], Pop the top off stk and
        go back to Step 3. /* recall nj is the numerical join function */
Step 4. /* top of stk is -1, signalling lp */
        Let n = n - 1; If n = 0 then return(v) and exit.
Step 5. Pop the top off stk and then Push v onto stk. Go to Step 2.

Step 6. /* we've reached end of bit string; append rps as needed */
        If n ≤ 0 or stk is empty then return(v) and exit.
Step 7. /* test for dangling lps */
        If Top[stk] < 0 then
            Pop top off stk; Let v = nj[v,0]; Let n = n - 1; go back to Step 6.
Step 8. Let v = nj[Top[stk],v]; Pop top off stk; go back to Step 6.
```

---

In case the bits are shifted into a register, here is another algorithm for decoding the bitstring, starting at the end of the string.

---

```
Step 0. /* Algorithm to decode a BSX codeword as a number, where the input bits are
processed serially starting with the last bit. Considered as the code for an
ordered bare tree, the nodes are transversed in post-order. */
      Input arguments:
       x; /*   x = (x[0], x[1], ..., x[# x - 1]) as a string of 0s and 1s. x is
assumed to be a valid binary BSX codeword. */
      Output value:
       v; /* decoded returned value */
      Local variables:
       m, /* string length of x */
       h, /* head popped off first */
       t, /* tail popped off second */
       p, /* index of current position in string */
       b, /* bit currently input from string x */
       stk; /* stack of computed values */
Step 1. Let m = #x; If m = 0 then return(0) and exit. Otherwise, Let p = m-2;
```

```
          /* a binary BSX string must end in a 1 */
        Let stk={0}.
Step 2. /* main loop to read bits */
        Let p = p - 1;If p < 0 then Go to Step 3.
        Otherwise, Let b = x[p].
        If b = 0 then /* found another lp */
            Pop h off stk;
             Pop t off stk;
   /* recall nj is the numerical join function */
              Push nj(h,t) onto stk.
        Otherwise, Push 0 onto stk; Go to Step 2.
Step 3. /* end of string */
    Let v = Top[stk]. Return(v) and exit.
```

---

The encoding of $x$ as a bsx, $X_E(x)$ results in a string of length $\#e(x) = 2\,\mathrm{lgx}(x) + 1$ , so adding the start sequence 01 the total transmission time is $(2\,\mathrm{lgx}(x) + 1)\delta$ . Since $\mathrm{lgx}(x) = \min\{n\in\mathbb{N} \mid x < S_{n+1}\}$, we need to study the rate of growth of $S_n$.

*Proposition (7.1). Let $C_n$ be the nth Catalan number and $S_n = C_0 + \ldots + C_{n-1}$ . Then $S_n \sim \frac{n^{-3/2}}{3\sqrt{\pi}}\, 4^n$ .*

*Proof:* The method of the proof is to show that $S_n \sim \frac{4}{3}\, C_{n-1}$ and then use Stirling's approximation to estimate the growth of $C_{n-1}$.

*Lemma (7.2) . $S_n \sim \frac{4}{3}\, C_{n-1}$*

Rewrite $S_n = \sum_{j\in n} C_{n-1-j}$ . Now in an appendix below it's established that the Catalan numbers satisfy

$C_i = (4 - 6/(i+1))\, C_{i-1}$ for all $i > 0$, and $C_0 = 1$. By induction $C_i = \prod_{k=2}^{i+1} (4 - 6/k)$,

therefore $C_{n-1-j}\big/C_{n-1} = \prod_{k=n-j+1}^{n} \left(\frac{1}{4}\quad k/(k-3/2)\right) = (1/4)^j \prod_{k=n-j+1}^{n} \left(1 + \frac{3}{2k-3}\right)$ Thus

$S_n = C_{n-1} \sum_{j=0}^{\infty} a_{nj}$ where $a_{nj} = [\![\, j < n ]\!]\ (1/4)^j \prod_{i=0}^{j-1} \left(1 + \frac{3}{2(n-j+1+i)}\right)$. For fixed $j$,

if $n > j$ then for each $i \in j$ the factor $\left(1 + \frac{3}{2(n-j+1+i)}\right)$ decreases to 1 as $n \to \infty$,

and so their product decreases to 1, for each $j$, and therefore $\lim_{n\to\infty} a_{nj} = (1/4)^j$.

On the other hand, if $n > j > i \geq 0$ then $\left(1 + \frac{3}{2(n-j+1+i)}\right) \geq (1 + 3/4)$ so for every $j$,

$a_{nj} \leq (1/4)^j\,(7/4)^j = (7/16)^j$, so by the dominated convergence theorem,

$\lim_{n\to\infty} \sum_{j=0}^{\infty} a_{nj} = \sum_{j=0}^{\infty} \lim_{n\to\infty} a_{nj} = \sum_{j=0}^{\infty} (1/4)^j = 1/(1-1/4) = 4/3$. This proves the lemma.

Now since $\frac{C_n}{C_{n-1}} = 4 - 6/(n+1)$, lemma (7.1) clearly implies $S_n \sim (1/3)\, C_n$ . The proposition will follow from the following lemma:

Lemma (7.3) $C_n = \frac{n}{n+1} \frac{e^{\theta_n}}{\sqrt{\pi}} n^{-3/2} 4^n$ , where $-1/(6n) < \theta_n < 0$.

*In particular,* $C_n \sim \frac{1}{\sqrt{\pi}} n^{-3/2} 4^n$

We can write $C_n = 1/(n+1) \frac{(2n)!}{n!\, n!}$ . One form of Stirling's approximation is: $m! = \sqrt{2\pi m}\; m^m e^{-m+\lambda_m}$ , where $1/(12m+1) < \lambda_m < 1/(12m)$. Then

$$C_n = 1/(n+1) \frac{\sqrt{4\pi n}\,(2n)^{2n} e^{-2n+\lambda_{2n}}}{2\pi n\, n^{2n} e^{-2n+2\lambda_n}} = \frac{\exp(\lambda_{2n}-2\lambda_n)}{n+1} (\pi n)^{-1/2} 2^{2n} = \frac{n}{n+1} \frac{e^{\theta_n}}{\sqrt{\pi}} n^{-3/2} 4^n, \text{ with}$$

$\theta_n = \lambda_{2n} - 2\lambda_n$, and $-1/(6n) < 1/(24n+1) - 2/(12n) < \theta_n < 1/(24n) - 2/(12n+1) < 0$.

*Proposition (7.2) .* $lgx(x) \sim \frac{1}{2} lg_2(x)$

*Proof:* By Prop. (7.1), $1 + S_n \sim 2^b n^{-3/2} 4^n$ , where $b = \log_2(1/(3\sqrt{\pi})$ ) is a constant, so

$\log_2(1 + S_n) = h_n + b - (3/2)\log_2(n) + 2n$, where $h_n \to 0$ as $n \to \infty$ . Upon dividing by $2n$ it is clear that $\log_2(1 + S_n)/(2n)$ converges to 1, albeit slowly. The conclusion holds when we apply the floor function to the $\log_2$ ; i.e.,

$\lg_2(S_n) \sim 2n$, and this remains true if $n$ is replaced by $\text{lgx}(x)$ . Now if $n = \text{lgx}(x)$ then $S_n \leq x < S_{n+1}$ so $\lg_2(S_n) \leq \lg_2(x) \leq \lg_2(S_{n+1})$, whence

$\lg_2(S_n)/(2n) \leq \lg_2(x)/(2n) \leq \lg_2(S_{n+1})/(2n+2) \frac{n+1}{n}$. As $x \to \infty$, $n = \text{lgx}(x) \to \infty$, and so the proposition follows.

*Corollary (7.1) The BSX encoding is asymptotically optimal for monotone prefix-free codes.*

$(2\,\text{lgx}(x) + 1)/\lg_2(x) \sim 1$ .

# 8. The Probability Distribution Determined by The BSX encoding and its Analytic Continuation

The characteristic sum of a binary code is defined by $\sigma = \sum_{x\in\mathbb{N}} 2^{-L_B(x)} = \sum_{k\in\mathbb{N}} \nu_k\; 2^{-\lambda_k}$ . According to the Kraft inequality [Bobrow and Arbib], $\sigma \leq 1$ for a prefix-free monotone code. $-\log_2(\sigma)$ is a measure of the redundancy of the code, so that when $\sigma = 1$ there is no redundancy; we show this to be the case for the bsx code. In any case, $x \mapsto 2^{-L_B(x)}/\sigma$ defines a probability distribution on $\mathbb{N}$, giving a message source structure on $\mathbb{N}$ . For non-redundant codes with $\sigma = 1$ , the source entropy actually equals the expected value of the codeword length $L_B$, instead of providing a lower bound. However, in the bsx case this source turns out to have infinite entropy (and so infinite average word length), but closely related source distributions show less extreme randomness.

First, let us note that for the bsx code $X$, $\nu_k = C_k$ and $\lambda_k = 2k + 1$ . Chaitin studied this probability distribution ; e.g., [AIT].

*The binary symbolic expression [bsx] distribution with parameter z* $\in$ ]*0,1/4*] is the probability distribution on $\mathbb{N}$ with weight function

$w_X(z)(x) = \frac{z^{lgx(x)}}{G(z)}$ , where $G(z)$ is the generating function (4) for the Catalan numbers.

$w_X(1/4)(x) = \frac{(1/4)^{lgx(x)}}{G(1/4)} = 2^{-L_X(x)}$, the source distribution associated with the bsx code.

First we note that the power series for $G(z)$ actually has radius of convergence 1/4, since $C_{n+1}/C_n \to 4$ . To show $w = w_X(z)$ defines a probability distribution, for $0 < z < 1/4$ we have

$\sum_{x\in\mathbb{N}} w(x) = \sum_{n=0}^{\infty} \sum_{\text{lgx}(x)=n} z^n/G(z) = \sum_{n=0}^{\infty} C_n\, z^n/G(z) = 1$ by definition of $G$. When $z = 1/4$ we could appeal to the Kraft inequal-ity to establish the convergence, or merely observe that lemma 4.2 implies that $C_n\ (1/4)^n = O\left(n^{-3/2}\right)$. Then, by Abel's theorem in series,

$\sum_{x\in\mathbb{N}} w(1/4)(x) = \sum_{n=0}^{\infty} C_n\ (1/4)^n / G(1/4) = \lim_{z\to 1/4} \sum_{n=0}^{\infty} C_n\ z^n = G(1/4) / G(1/4) = 1.$

Note Pr({0}) = 1/G(z), and this ranges from 1/2 when z=1/4 to 1 as z → 0. The event {0} corresponds to a bsx chosen at random (according to the distribution induced from $w_z$ ) being equal to nil, "()" with codeword "1" . If we let $p_0$= Pr({0}) then the equation satisfied by $G(z)$ implies $z = (1 - p_0)\ p_0$ .

Since $w(z)(x)$ depends on $x$ only through lgx($x$), it is convenient in the rest of this section to use $N(x)$ for lgx($x$) and refer to $N(x)$ as the size of $x$. We know $N^{-1}[\{n\}] = [\ S_n,\ S_{n+1}[$ and $\#N^{-1}[\{n\}] = C_n$. It follows immediately that

$$\Pr(N = n)\ =\ C_n\, z^n\ /\, G(z) \qquad (8)$$

We can decompose the probability according to the conditional probability induced by $N$, and writing Ex for expected value, we have

$\mathrm{Ex}\,(F)\ =\ \sum_{x\in\mathbb{N}} F(x)\, w(z)\,(x)\ =\ \sum_{n\geq 0} \Pr(N = n)\ \left((1/C_n) \sum_{x:N(x)=n} F(x)\right)$, and the conditional probability measure Pr( | $N = n$) with support in $N^{-1}[\{n\}]$ is distributed uniformly over the $C_n$ members of that set.

This suggests that to simulate a random selection $X$ from the distribution determined by $w_X$ we can first select $N$ according to the distribution (4), and then choose an integer $R$ uniformly from the interval $[0,\ C_n - 1]$, and let $X\ =\ S_N\ +\ R$.

- **Example: The mean of the numerical value, Ex($X$)**

Let's apply these considerations for the random variable $X$ , where $X(x) = x$ . Since we are summing an arithmetic progression, the average of $X$ over $N^{-1}[\{n\}]$ is $(S_n + S_{n+1} - 1)/2\ = S_n + (C_n - 1)/2$ . Therefore

$$\mathrm{Ex}(X)\ =\ \sum_{x\in\mathbb{N}} x\ w(z)\,(x)\ = \sum_{n\in\mathbb{N}} \frac{C_n}{G(z)}\, z^n\,(S_n\ +\ 1/2\, C_n\ - 1/2)\ =\ \sum_{n\in\mathbb{N}} \frac{z^n}{G(z)}\, C_n\,(S_n\ +\ 1/2\, C_n)\ - 1/2$$

Applying the results of lemmas (7.1 & 2) , we see that $C_n\ (S_n + 1/2\ C_n) \sim 5/6\ \ C_n^{\ 2}\ \sim (1/\pi)\ n^{-3}\ 16^n$, so that the infinite series for Ex($X$) converges if 0 < z ≤ 16 and diverges to infinity otherwise . Thus as a function of $z$, Ex($X$) increases to a maximum of

$$\mathrm{Ex}(X)_{\max}\ =\ \sum_{n\in\mathbb{N}} \frac{(1/16)^n}{G(1/16)}\, C_n\,(S_n\ +\ 1/2\, C_n)\ - 1/2\ \ = 0.0916\ .\ \text{when}\ z\ =\ 1/16\ .$$

For z ∈ ]1/16, 1/4] , Ex($X$) jumps to infinity. Note that when z=1/16, $\Pr(\{0\}) = 1/\left(8\left(1 - \sqrt{3}\ /2\right)\right) = 0.9330$, so when sampling from this distribution, the probability of 0 is well over ninety percent.

- **The mean and entropy of the distribution of *N***

We are interested in quantities more related to the structure transported to ℕ by the $d/e$ encoding of section 6. Recall that

$\sharp\, e(x)\ =\ 2\, N(x)\ +\ 2$ , and $\sharp\ X_E(x)\ =\ 2\, N(x)\ +\ 1$ .

$N(x)$ is also one less than the number of nodes in the rooted tree associated with the bsx $e(x)$ .

  - ***Case 0 < z < 1/4***

We can compute Ex($N$) by means of generating functions, and we get

$$\mathrm{Ex}(N)\ =\ (1/G(z)) \sum_{n\in\mathbb{N}} n\, C_n\, z^n\ =\ (1/G(z))\, z\, \frac{d}{dz}\, G(z)\ =\ \left(\frac{1}{\sqrt{1-4\,z}} - 1\right)\Big/ 2\,,\ \text{so}$$

$$\nu = \mathrm{Ex}(N) = \left(\frac{1}{\sqrt{1-4z}} - 1\right) \Big/ 2\,, \text{ and } z \text{ is determined by } \nu \text{ via } \quad z = \frac{\nu^2+\nu}{(2\,\nu+1)^2} \qquad (9)$$

The entropy of the distribution of $N$ is the sum $\sum_{n=0}^{\infty} -\log_2(\Pr(N{=}n))\,\Pr(N{=}n)$, and again by using the estimates of lemmas (7.1 & 2) it is easy to establish the convergence of the series when z < 1/4 , but we have been unable to obtain a closed form for the sum.

Plot of the entropy of the distribution of N as a function of z

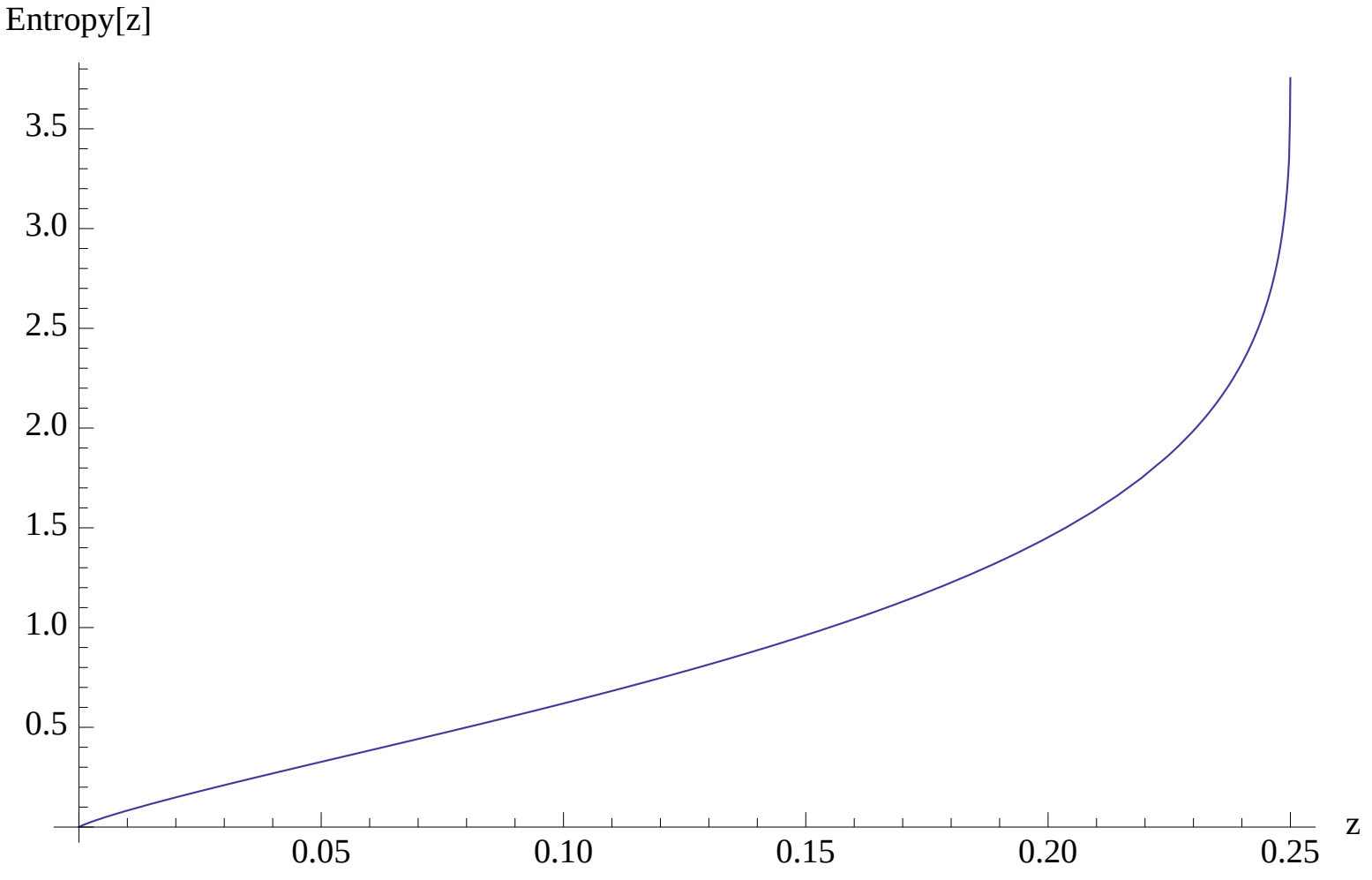

The entropy increases slowly to infinity as z approaches 1/4 .

- ***Case z = 1/4***

When z = 1/4 the distribution $w(1/4)$ on ℕ is the same as that determined by the bsx binary code, and $L_X = 2\,N + 1$ . In this case $\Pr(\{0\}) = w(1/4)\,(0) = 1/G(1/4) = 1/2$, so a bsx chosen at random has an even chance of being different from nil.

For this distribution,

$\mathrm{Ex}(N) = (1/2) \sum_{n\in\mathbb{N}} n\,C_n\,(1/4)^n$ . Using Lemma (7.2) we can say for $n > 0$,

$$n\,C_n\,(1/4)^n = \frac{n}{\sqrt{\pi}}\, e^{\theta_n} \left(\frac{n}{n+1}\right) n^{-3/2} > \frac{e^{-1/6}}{2\sqrt{\pi}}\; n^{-1/2}$$

and of course the series

$\sum_{n=1}^{\infty} n^{-1/2}$ diverges to infinity. $A$ similar argument demonstrates that the entropy of the $w(1/4)$ distribution is infinite.

As mentioned earlier, $w(1/4)$ is identical to the distribution defined by the characteristic sum: $w(1/4)(x) = 2^{-L_X(x)}$ . As a binary code, this code has an infinite mean word length. However, the analytic continuation $w(z)$ for $0 < z < 1/4$ does have finite mean size and finite entropy, and in fact z may be determined by specifying any positive average length. And for all of them, certain statistical relations among the numerical versions of joining and separating stay the same.

## The Distribution of Numerical Join, Head, and Tail

In the following, the probability space is ℕ with the measure Pr determined by the weights $w(z)$ for a fixed $z \in\, ]0,1/4]$.

First, we state as a proposition:

*Proposition (8.1) The size function N has the distribution: Pr(N=n) =* $\frac{C_n z^n}{G(z)}$.

$\nu = Ex(N) = \left(\frac{1}{\sqrt{1-4z}} - 1\right)/2$ *if* $z < 1/4$ *, and* $z = \frac{\nu^2 + \nu}{(2\nu + 1)^2}$ .

*Proof:* This was established as equation (8) above.

*Proposition (8.2) The join function J, considered as a random variable on* ℕ◊ℕ *with the product measure has the distribution:*

$\forall\, a \in \mathbb{Z}_+ = \mathbb{N}\sim\{0\},\ Pr(J=a) = w(z)(a) / Pr(\mathbb{Z}_+) = z^{N(a)}/(G(z)-1)$

*Proof:* Recall $J: \mathbb{N}\Diamond\mathbb{N} \to \mathbb{Z}_+$ where $J(x, y) = nj(x,y)$ and $nj$ the numerical join defined in Section 5. Given $a > 0$,

$$\Pr(J = a) = \sum_{(x,y):J(x,y)=a} w(z)(x)\ w(z)(y) = w(z)(nh(a))\, w(z)(nt(a)) = 1\big/\left(G(z)^2\right) z^{N(nh(a))+N(nt(a))}$$

$$= 1\big/\left(G(z)^2\right) z^{N(a)-1} = w(z)(a)/(zG(z))\,.\ \text{But } z\,G(z) = 1 - 1/G(z) = \Pr(\mathbb{N}\sim\{0\}) = \Pr(\mathbb{Z}_+)$$

The map $J$ is one-to-one and onto, and the proposition states that the probability measure that $J$ transports from ℕ◊ℕ to $\mathbb{Z}_+$ is just the conditional probability measure $\Pr(\ |\,\mathbb{Z}_+)$ on $\mathbb{Z}_+$ that it inherits from Pr .

*Proposition (8.3) Let* $H = nh\,|_{\mathbb{Z}_+}$ = numeric head restricted to the positive integers, and *let* $T = nt\,|_{\mathbb{Z}_+}$. *Then considered as ℕ-valued random variables on the probability space* $(\mathbb{Z}_+, Pr(\ |\,\mathbb{Z}_+))$*, H and T are independent identically distributed random variables, with common distribution Pr,* (weight function $w(z)$ ).

*Proof:* For $a > 0$, $J^{-1}(a) = (H(a), T(a))$, and the preceding proposition establishes that $J^{-1}$ transports $\Pr(\ |\,\mathbb{Z}_+)$ to Pr⊗Pr, the product measure on ℕ◊ℕ, which is to say that the components $H$ and $T$ are independent, each with distribution Pr.

Recall that a bsx may be interpreted as a (possibly empty) list of bsxes. Let us define the list length $M(x)$ to be the number of items on the list $e(x)$, so that $M: \mathbb{N} \xrightarrow{\text{onto}} \mathbb{N}$ . More formally,

*The list length M is defined recursively by M(0) = 0, and for x > 0,* $M(x) = 1 + M(T(x))$, where $T$ is the numeric tail function.

Since $T(x) < x$ if $x > 0$, $M$ is indeed well-defined.

*Proposition (8.4) The random variable M has a geometric distribution with parameter* $p_0$ = *Pr*({0})*=1/G(z).*

*Thus* $p_0 \geq 1/2$ . If $\mu = \text{Ex}(M)$ then $\mu = (1 - p_0)/p_0 \leq 1$, and $z = \mu\big/(1+\mu)^2$ .

*Proof:* $M(x) = 0$ iff $x = 0$, so $\Pr(M=0) = \Pr(\{0\}) = 1/G(z) = p_0$. So let $m > 0$ . Then

$$\Pr(M = m) = \sum_{x>0:\,M(x)=m} w(z)(x) = \sum_{s\geq 0}\sum_{t:\,M(t)=m-1} w(z)(J(s,t)) = \sum_{s\geq 0}\sum_{t:\,M(t)=m-1} w(z)(J(s,t))$$

$$= \sum_{s\geq 0}\sum_{t:\,M(t)=m-1} (1/G(z))\, z^{N(s)+N(t)+1} = \left(\sum_{s\geq 0} w(z)(s)\right) z\,G(z) \sum_{t:\,M(t)=m-1} w(z)(t)$$

$$= z\,G(z)\Pr(M = m-1)\,.\ \text{The solution to this recursive equation is } \Pr(M = m) = \Pr(M = 0)\cdot(z\,G(z))^m$$

But $z\,G(z) = 1 - 1/G(z) = 1 - p_0$, so $\Pr(M = m) = p_0 \cdot (1 - p_0)^m$ . The mean of such a distribution is $\mu = (1 - p_0)/p_0 = G(z) - 1$ , whence the equation for $z$ in terms of $\mu$ .

Now for each $m \in \mathbb{N}$ we shall consider the subspace

$\mathbb{L}_m = M^{-1}[\{m\}] = \{x \in \mathbb{N} \mid \exists\, \epsilon \in BSX_m$ *such that e(x) = "(" <> ϵ*$_0$ *... <> ϵ*$_{m-1}$*<> ")"}* with the conditional probability measure

$A \longmapsto \Pr(A \mid M = m) = \Pr(A \cap \mathbb{L}_m)/\Pr(\mathbb{L}_m) = (1/p_0)\cdot(1 - p_0)^{-m} \sum_{a\in A} w(z)(a)$ . These measures have disjoint supports for distinct values of $m$, and Pr is the convex combination of the measures $\Pr(\ |M=m)$ :

$$\Pr = \sum_{m\in\mathbb{N}} p_0\cdot(1 - p_0)^m \Pr(\ \mid M = m) \qquad (10)$$

## The Fixed Point Property of this Family of Distributions

Next, for each $m$ consider the map $P^{(m)}: \mathbb{L}_m \xrightarrow{1-1\text{ onto}} \mathbb{N}^m$, where for $y \in \mathbb{L}_m$

$P^{(0)}(0) = \Lambda$ (empty list), *and for* m > 0, $P^{(m)}(y) = \left(P^{(m)}(y)_j\right)_{j \in m}$, $P^{(m)}(y)_j = H \circ T^{\circ j}(y) = H(T(...T(y)...))$,

where the numeric tail operation $T$ is applied $j$ times.

*Lemma:* The map $P^{(m)} : \mathbb{L}_m \overset{\square}{\to} \mathbb{N}^m$ *is one-to-one and onto the set of all lists of natural numbers of length m . The inverse map is given by* $P^{(m)-1}(\xi) = J(\xi_0, J(\xi_1, ..., J(\xi_{m-1}, 0)...))$

*Proof:* Let $Q^{(m)}(\xi) = J(\xi_0, J(\xi_1, ..., J(\xi_{m-1}, 0)...))$. Then verify by direction substitution, that $\forall \xi \in \mathbb{N}^m$, $P^{(m)}(Q^{(m)}(\xi)) = \xi$ and $\forall x \in \mathbb{L}_m$, $Q^{(m)}(P^m(x)) = x$ .

Now the space $\mathbb{N}^m$ with the product measure $\Pr^{\otimes m}$ may be considered as the space of random samples of size $m$ from the distribution Pr. Its weight function is ${w_X}^{\otimes m}$, where

$$ {w_X}^{\otimes m}(\xi) = \prod_{j=0}^{m-1} w_X(\xi_j) = G(z)^{-m} z^{\sum_{j=0}^{m-1} \mathrm{lgx}(\xi_j)} \tag{11}$$

*Proposition (8.5) The conditional probability Pr( |M=m) is the distribution of* $Q^{(m)}$ *as a r. v. on* $(\mathbb{N}^m, Pr^{\otimes m})$.

Proof: In terms of weights we must show that

$$\frac{w_X(Q^{(m)}(\xi))}{p_0 \cdot (1-p_0)^m} = {w_X}^{\otimes m}(\xi)$$

Recall

$$z \cdot G(z) = 1 - 1/G(z) = 1 - p_0 . \ \mathrm{lgx}(J(x, y)) = \mathrm{lgx}(x) + \mathrm{lgx}(y) + 1$$

Let $x = Q_{(m)}(\xi) = J(\xi_0, J(\xi_1, ..., J(\xi_{m-1}, 0)...))$. Then

$$\mathrm{lgx}(x) = m + \sum_{j=0}^{m-1} \mathrm{lgx}(\xi_j)$$

The LHS of the weight equation then reduces to

$$(z \cdot G(z)/(1 - 1/G(z)))^m \cdot \prod_{j=0}^{m-1} w_X(\xi_j) = {w_X}^{\otimes m}(\xi)$$

using the identity preceding.

One way of phrasing this proposition is that the $\mathbb{N}$-valued r.v.s ${P^{(m)}}_0, {P^{(m)}}_1, ..., {P^{(m)}}_{m-1}$ on Pr( |M=$m$) are $m$ i.i.d. r.v.s with Pr as the common distribution (recall Pr is the probability measure with weight function $w_X$).

*Theorem(8.1): Let* $P^{(*)} = \bigcup_m P^{(m)}$, *so that* $P^{(*)}: \mathbb{N} \xrightarrow{1-1\text{ onto}} \mathbb{N}^*$ *and* $P^{(*)}(x)_j = H \circ T^{\circ j}(x)$ *for* $0 \le j < M(x)$ *. Then in term of probability weights,*

$w_X\left(P^{(*)-1}(\xi)\right) = p_0 (1-p_0)^{\sharp \xi} {w_X}^{\otimes \sharp \xi}(\xi) = (p_0{}^{\sharp \xi}) \cdot z^{\sum_j lgx(\xi_j)}$ *. In terms of probability measures,*

$\left(P^{(*)}\right)_*(Pr) = \oplus_m p_0 \cdot (1 - p_0)^m Pr^{\otimes m}$ .

*Proof*: The result about probability weights follows from the definition of $P^{(*)}$, which when restricted to $\mathbb{L}_m$ equals $P^{(m)}$, and proposition (8.5). The expression in terms of Pr, the probability measure associated with weight $w_X$, is a consequence of the definitions in the appendix together with equation (10) and proposition (8.5).

The formulation in terms of probability measures is curious: it says that Pr is isomorphic to its sampling space, represented by the RHS of the equation. It is essentially a fixed point of the transformation which takes a probability measure to its sampling space, at least if the sample size has a geometric distribution with mean at most 1 .

# 9. An Effective Recursive Enumeration of All Partial Recursive Functions

Since the 1930's there certainly have been computable recursive enumerations of the set of all partial recursive function; it was critical to the work of Kleene, Church et. al. to prove that for each positive integer $n$ there exists a computable map $\varphi^{(n)}$: $\mathbb{N} \overset{\text{onto}}{\longrightarrow} \mathcal{R}_n$, the set of all partial recursive functions whose domain is a subset of $\mathbb{N}^n$ and whose range a subset of $\mathbb{N}$ . The map $(x,y) \longrightarrow (\varphi^{(n)})_x(y)$ is required to be a partial recursive function of $n$+1 variables. If $f$ is a partial recursive function of n variables, then a natural number $x$ such that $f = (\varphi^{(n)})_x$ is called a Gödel number for the function $f$. A given function will always have infinitely many Gödel numbers.

The emphasis in the early days was to convince the community that such a computation, whether for Turing machines, lambda expressions, or whatever, was possible, even straightforward, but dreadfully tedious. Computational models implemented on computers were rare.
With the advent of Lisp in 1958, scholars could actually program and experiment with such operations. Mostly, they were based using the form of the expression Evaluate[Apply[($\lambda v.b$), '$e(y)$]] in an empty environment, where here $e$ represents whatever encoding of the natural numbers used by the version of the lambda calculus. '$e(y)$ represents an expression whose value is always $e(y)$. If we assume $n = 1$ we may assume the variable $v$ is fixed. The value of the function defined is $d(z)$, where $z$ is the value of $b$ in an environment wherein the variable $v$ has the value $e(y)$, and $d$ is the suitable decoder from expressions to natural numbers. Now by Church's thesis, this maps all lambda expressions of the variable $v$ onto the set of all partial recursive functions of one variable, $y$ . By our conventions, the lambda expression is completely determined by the function body $b$, which might intentionally contain $v$ as a free variable. So, we can construct a Gödel numbering of the partial recursive functions of one variable, provided we can construct a function $p$ from $\mathbb{N}$ onto all the expressions in our language (without any restrictions on $b$) .

In principle, the construction of such a function $p$ is straightforward: Fix a finite ordered alphabet $A$ (such as the ASCII character set) so that our language (the lambda calculus, say) is a subset of $A^+$. Given $x$, start through the non-empty strings on $A$ in lexicographic order, incrementing a counter each time the string is an expression in our language. Stop when the counter reaches $x$ and the corresponding string is $p(x)$ . With $p$, we can define $\varphi_x(y) = d$(Evaluate[Apply[($\lambda v.p(x)$), '$e(y)$]]) , defined for all $y$ such that the application evaluation terminates.

The brute force enumeration just described has the property that the stringlength of $p(x)$ is non-decreasing with $x$, so the larger $x$ is presumably the longer and more complicated is the program. Given a Gödel numbering $\varphi$ of functions of one variable, we could define a measure of complexity of a given partial recursive function $f$ by $M(\varphi)(f) = min \{x \mid \varphi_x = f\}$, the smallest Gödel number for $f$ .

The Kolmogorov complexity of $f$ (relative to $\varphi$) is essentially $K(\varphi)(f) = \lfloor \log_2(1+M(\varphi)(f)) \rfloor$ , the number of bits to specify $f$ . But to store the data to construct $f$ we must also store $K$ in addition to the bits themselves; in order to communicate the data over a binary channel we must encode then data so we can detect the end of transmission. We are thus led to the considerations of the preceding section; that is, we seek prefix-free binary codes for the natural numbers. Then we could modify the definition, given a prefix-free code ($B_E$ , $B_D$) as well as $\varphi$, so that the Gödel-Chaitin complexity of $f$ is the smallest length of all the codewords which encode a Gödel number for $f$ . Thus

$$G(\varphi)(f) = min \{\#s \mid s \in (2)^+ \text{ and } \varphi_{B_D(s)} = f\}$$

Two contemporary investigators in this line of development, Gregory Chaitin [http://www.cs.auckland.ac.nz/CDMTCS/chaitin/] and John Tromp [http://en.wikipedia.org/wiki/Binary_Lambda_Calculus] bypass Gödel numbers and work respectively with Chaitin's version of Lisp easily encoded in binary, and a binary lambda calculus developed by Tromp. Both these implementations have the advantage of being able to define complexity concretely and prove theorems about it. The lambda calculus and all its derivatives (including BILL described in the next subsection) are self-limiting basically because of the requirements that parentheses be nested. One can prove theorems that certain procedures exist by actually writing programs which implement them.

## BILL

The Lisp programming language is built around forming lists by enclosing items in parentheses. According to some programmers, Lisp stands for Lost In Stupid Parentheses; such programmer will find BILL the ultimate swamp of () pairs. Nevertheless, the language is of universal power equivalent to the Church lambda calculus.
Because the strings of balanced () pairs can be enumerated efficiently, BILL provides a way for the enumeration of algorithms alternative to an enumeration of Turing machines.

Like Lisp itself, BILL is based on the evaluation of symbolic expressions; the difference is that BILL has only two characters, '(' and ')', with which to form symbolic expressions. Thus the syntax of BILL was described in Section (2), and the expressions of the language are merely the bsxes. The question now is: How does BILL evaluate bsxes?

### Variables and Constants

A more basic question is how does BILL handle variables and determine their value? Well, variable expressions are of the form $v = (a)$, where $a$ is a bsx $\neq$ *nil* . The value of a variable expression is always relative to an *environment*. The variables are bound to specific values by an environment. For our purposes, an environment is a bsx interpreted as a look-up table, with an entry followed by its value, so conceptually the list is divided into pairs. So to evaluate ($a$) in the environment $e$, starting with *head*($e$), we search for the first occurrence of a pair of items starting with $a$, and then the value is the following item. If $a$ does not occur in an odd position on the list $e$ then the value of ($a$) is just $a$ itself. For example, suppose

$e$ = ( (())((())) (()())(()()()) (())(()()()) (()()())() ). The value of ((())) is ((())) [not (()()()) ] . The value of ((()())) is (()()()) . The value of (((()))) is ((())) because ((())) doesn't occur in an odd position. Finally, the value of ((()()())) is () .

Obviously if the variables in BILL are to be readable, we need a way to get from non-empty strings of letters to bsxes. The idea is to define a one-to-one map, $n$ say, from non-empty strings of letters onto positive integers. Well, let $b$ (the base) be an integer > 1. A standard one-to-one mapping from the strings on the alphabet $b$ = {0,1,...,$b$-1} onto the natural numbers:

$$n(\xi) = (b^n - 1)/(b-1) + \sum_{k\in n} \xi_k\, b^k\ ,\ \xi \in (b)^n = \text{strings of length } n \text{ on the alphabet } b\,.$$

We apply the above to the case where $b$ = 52 and use the string

$L$ = "abcdefghijklmnopqrstuvwxyzABCDEFGHIJKLMNOPQRSTUVWXYZ"

to set up a one-to-one correspondence between the ASCII lower and upper case letters and the alphabet 52. That is, $L_0$ = 'a' and $L_{51}$ = 'Z' . Next, we follow this map with the bsx encoder $e$ described above, and we have a simple algorithm for the function $\tilde{L}$ converting variable names into bsxes. Thus the variable name $a$ translates as "(())", and the variable ($a$) translates as ((())) = *join* ($\tilde{L}$("a"),"()"), $b$ as ((()())), and the variable named *variable* as "((((((()((()()(()))))(()())(()()))(()()()()((()))()))))", as it turns out.

In this way we see that BILL can have an infinite number of variable expressions. To change the value of the variable we have to append the desired value and then the variable to the current environment. The expressions () and (()) are constants whose value is always itself in any environment. () and (()) play the role of False and True in BILL.

The following recursive algorithm demonstrates how the value of a variable is determined by an environment:

---

```
Function lookupValue[v,e]
Step 0. /* Function which looks up value of a variable */
      Input Arguments:
      v /* variable label whose value is sought */,
      e /* environment */;
      Output Returned Value of Function:
      a /* new environment built by assigning evaluated args to variables */;
      Local variables:
      a /* value of next argument  */;
```

```
Step 1. /* Test for empty environment */
    If e = nil then Return[v]; /* ...since v was not found. */
Step 2. /*  */
    If v = head[e] then Return[head[tail[e]]]; /* v was found */
Step 3. /* look up in rest of list  */
    Return[lookupValue[v,tail[tail[e]]]]; /* recursive call *
```

---

By means of the encoding *e* described in Theorem (1), we can generate constant expressions corresponding to any natural number in ordinary decimal notation. To this end we need something like the Quote operation of standard Lisp, which copies its argument without evaluating it. In BILL a quote operation is implemented by joining "()" to a non-*nil* string; the quote of the constants () and (()) are themselves. So to represent the number constant 5 , we encode 5 as "(()(()))" and build the expression "(()()(()))" , whose value in any environment will be "(()(()))" . The constant expressions corresponding to 0 and 1 are () and (()) respectively. The constant "(())" is named T [analogous to Lisp].

The decimal strings "0", "00", "000", "0000",...may be used to represent the special constants $S_0$=0, $S_1$=1, $S_2$=2, $S_3$=4..., which would be encoded (using quotes after the first two) as the bsxes "()", "(())", "(()()())", "(()()()())", ...

## Primitive Functions

The current version of BILL has only five primitive functions. Missing arguments default to *nil*, and extra arguments are always ignored. Like variables, primitive functions are encoded as lists with one item

### *if*

*if* is a function which takes three arguments, *if*[*a*,*b*,*c*]. *a* is evaluated in the current environment. If the value of *a* is not *nil* then *b* is evaluated, and its value is that of *if*[*a*,*b*,*c*]. But if the value of *a* is *nil* then *b* is not evaluated; *c* is evaluated instead, and its value is the value of *if*[*a*,*b*,*c*] . *if* is how BILL does selection; it is descended from Lisp's COND function.

The function code for *if* is 0, and its expression is (()) .

The next three functions are the functions *join, head,* and *tail* but with their arguments evaluated first. Were we to call them *push, top,* and *pop* we would emphasize that to a computer scientist our bsxes are stacks of stacks, starting with the empty stack.

### *join*

*join*[*a*,*b*] first evaluates *a* and *b* to get *x* and *y* respectively, and then its value is *join*(*x*,*y*) .

The function code for join is 1, and its expression is ((())) . It is descended from CONS .

### *head*

*head*[*a*] first evaluates *a* to get *x*, and then its value is *head*(*x*) .

The function code for head is 2, and its expression is ((()())) . It is analogous to CAR .

### *tail*

tail[*a*] first evaluates *a* to get *x*, and then its value is *tail*(*x*) . It is like CDR in Lisp.

The function code for tail is 3, and its expression is (((()))) .

### *out*

The function out is not essential, but it is a way for generating intermediate output as a side effect. out[*a*] has the value *x*, which is the value of the argument *a*. As a side effect, *x* is output to a earlier specified device in a chosen format.

The function code for out is 4, and its expression is ((()()())) . It is very helpful for debugging.

## Defined Functions

What makes a programming language like BILL so powerful, in spite of its limited primitive operations, is its ability to define arbitrary functions in a simple syntax, the lambda calculus without lambda. Defined functions are expressions $f = join(b,v)$, where $b = head(f)$, is the body of the function and $v = tail(f)$ is the function variable list, where $v \neq nil$ (this is how defined functions are distinguished from primitive functions).

When a defined function $f$ is evaluated given a list $t \neq nil$ of arguments a new environment is built on top of the current one, by joining to it the value of an argument and then joining the variable on the top of the variable list. This continues until the variable list is *nil*.

---

```
Function passArgs[t,v,e]
Step 0. /* Function which passes arguments */
      Input Arguments:
      t /* list of unevaluated bsxes */,
      v /* list of bsxes representing variable names */,
      e /* initial environment */;
      Output Returned Value of Function:
      n /* new environment built by assigning evaluated args to variables */;
      Local variables:
      a /* value of next argument  */;
Step 1. /* Initialize */
    Let n = e;
Step 2. /* Main loop */
    If v = nil then Return[n]; /* all args are passed */
Step 3. /* evaluate and pass next arg */
    Let a = bsxEval[head[t],e]; /* note evaluation takes place in e, not n ! */
    Let n = join[head[v],join[a,n]];  /* pass it! */
Step 4. /* Update and go back to test */
    Let t = tail[t];
    Let v = tail[v];
    Go to Step 2.
```

---

## The Function bsxEval

We need to describe more carefully the algorithm that evaluates a bsx $s$ in an environment $e$:

---

```
Function bsxEval[s,e]
Step 0. /* Function which evaulates bsxes in a given environment */
/* Since not all expressions have a value, this procedure can go crash! */
      Input Arguments:
      s /* bsx to be evaluated */,
      e /* environment table with values of variables */;
      Output Returned Value of Function:
      The value of s in e.
      Local variables:
      h /* head of s */,
      t /* tail of s, argument list */,
      f /* function header, value of h in e */,
      n /* new environment built by assigning evaluated args to variables */,
      b /* body of defined function */,
      v /* variable list */
      a /* value of first argument  */;
Step 1. /* Test for constants */
    If s = nil then Return[nil];
    If s = T then Return[T];
Step 2. /* separate into head and tail */
```

```
    Let h = head[s]; Let t = tail[s];
     If t = nil then Return[lookupValue[h,e]]; /* s is a variable, return its value
in e */
Step 3. /* Evaluate h to get function header f */
    Let f = bsxEval[h,e]; /* first recursive call */
    If f = nil then Return[t]; /* s is probably 't */
Step 4. /* separate f into body and variable list */
/* OK, t ≠ nil, so t represents a list of unevaulated arguments */
    Let b = head[f];
    Let v = tail[f];
    If v = nil then Go to Step 6. /* primitive function */
Step 5. /* Defined function */
    Let n = passArgs[t,v,e]; /* build new environment; resursive call */
    Return[bsxEval[b,n]];
Step 6. /* primitive function */
    Let a = bsxEval[head[t]];  /*  evaluate 1st arg  */
/* OK, v = nil and f = (b) . b is the code for the primitive function */
    If b = nil then Go to Step 7; /* if */
    If b = "(())"  then Return[join[a,bsxEval[head[tail[t]],e]]];
    If b = "(()())" then Return[head[a]];
    If b = "((()))" then Return[tail[a]];
    If b = "(()()())" then Output[a]; /* whatever that means... */
    Return[a] /* default */
Step 7. /* if */
    If a = nil then Return[bsxEval[head[tail[tail[t]]],e]]; /* condition failed */
    Return[bsxEval[head[tail[t]],e]]    /* condition satisfied */
```

---

This brief note is not a complete exposition of BILL, although we hope we have communicated the essentials. To show BILL is Turing complete it suffices to show that BILL can emulate the lambda calculus. We will only sketch what needs to be done and point to links in the appendix to *Mathematica* programs which implement some of the mentioned procedures.

The lambda calculus abstraction ($\lambda v.b$) corresponds to *join*[*b*,*v*] where *b* and *v* are bsxes, *v* used as a list of variable names and *b* as the body of the function. The result of applying the defined function *join*[*b*,*v*] to a list of arguments *t* is the bsx

*join*[‘*join*[*b*,*v*],*t*] = *join*[*join*[(),*join*[*b*,*v*]],*t*] . When this expression is evaluated by bsxEval in the *nil* environment, we get *join*[*nil*,*join*[*b*,*v*]]=’*join*[*b*,*v*] if *t* is the empty list (). Otherwise, the function header is *join*[*b*,*v*], which has head *b* and tail *v*.

If *v* is *nil*, we evaluate the primitive function encoded by *b* using the values of the arguments on t. If *v*≠*nil* then we have a defined function, and we evaluate *b* in an environment wherein the items on *v* are paired off with the value(in *nil* environment) of the corresponding item on *t*.

## A Gödel number mapping in BILL

$$\forall n,x\in\mathbb{N},\ let(\varphi^{(n)})_x(y) = d[bsxEval[('e[x]\ 'e[y_0]\ \ldots 'e[y_{n-1}]),nil]],$$

defined $\forall\ y \in \mathbb{N}^n$ such that the bsxEval expression on the RHS halts returning a bsx value which is then decoded. Note

$$\left(\varphi^{(0)}\right)_x(\Lambda) = d['e[x]] = d[join[nil,\ e[x]]] = nj[0,\ x] = C_{lgx[x]}+x$$

$(\varphi^{(n)})_0(y) = d[('e[y_0]\ \ldots 'e[y_{n-1}]]$ decodes a list of quoted arguments.

Church’s thesis implies that for each *n* the map $x \mapsto (\varphi^{(n)})_x$ takes ℕ <u>onto</u> the set of <u>all</u> partial recursive [i.e., computable] ℕ-valued functions of *n* ℕ-valued variables. Given a computable function *f* of arity *n*, to get a Gödel number *x* for *f*, write a program in the BILL language which implements *f*, producing $p \in BSX_*$, and let $x = d[p]$ . The determination of Gödel numbers using BILL is explicitly associated with the problem of producing programs in BILL, which can be done. We emphasize that any bsx can serve to define a function of any number of variables; there is no special syntax.

### ■ Arithmetic in BILL

In most presentations of the lambda calculus arithmetic is introduced by Church numerals. In BILL this might take the form

0::=((x)fx)

1::=((fx)fx)

2::=((f(fx))fx), etc.

where *f* is "((()))" and *x* is "((()()))" .

But in BILL we can identify each and every bsx with a unique natural number, using the encoding and decoding presented in Section 6. We can take the easy way out and define a successor function by $f(x) = e(d(x)+1)$, but let us consider an algorithm for a successor function which is defined for all bsxes; it is easily expressed in BILL (but not now!).

bsxes are ordered first by size and then by comparing the heads, and if they are equal comparing the tails. For a fixed size *n*, the first is "$(()\overbrace{\ldots}^{n}())$" and the last is "$((\overbrace{\ldots}^{n}()\overbrace{\ldots)}^{n})$" . The successor to the latter is "$(\overbrace{()\ldots()}^{n+1})$"; the size wraps around to $n+1$ . So the first step in writing a successor function is to write a function which detects bsxes of the form "$((\overbrace{\ldots}^{n}()\overbrace{\ldots)}^{n})$" and if of that form, computes the successor "$(\overbrace{()\ldots()}^{n+1})$" . As a consequence of Proposition 3, $e(S_n) =$ "$(()\overbrace{\ldots}^{n}())$" and $e(S_{n+1} - 1) =$ "$((\overbrace{\ldots}^{n}()\overbrace{\ldots)}^{n})$"

---

```
    Function up[x]
Step 0. /* Function up[x] */
      Input Arguments:
      x /* bsx to be tested for wrap-around. */
      Output Returned Value of Function:
      If no wrap-around occurs the output is nil.
      Otherwise, the output is succ[x] .
      Local variables:
      h /* up applied to head of x */
      t /* tail of x */
Step 1. /* Test for nil and T */
    If x = nil then Return[T]; /* () to (()) */
    If x = T then Return[join[nil,T]]; /* (()) to (()()) */
Step 2. /* Separate and analyze */
    Let t = tail[x];
    If t ≠ nil then Return[nil]; /* no wrap-around */
Step 3. /* Know t = nil. Test head[x] by recursive call */
    Let h = up[head[x]];
    If h = nil then Return[nil];  /* no wrap-around */
Step 4. /* wrap-around! */
    Return[join[nil,h]];
```

---

```
Step 0. /* Function which computes the next bsx after x */
      Input Arguments:
      x /* bsx whose successor is calculated */,
      Output Returned Value of Function:
      The value of e[d[x]+1] .
      Local variables:
      h /* head of x */,
      t /* tail of x */,
      u,v /* over-loaded and overworked temps */;
Step 1. /* Eliminate the easy case */
    If x = nil then Return T; /* () to (()) */
Step 2. /* test for wrap-around */
```

```
    Let v = up[x]; If v ≠ nil then Return[v];
Step 3. /* Separate into head and tail */
    Let h = head[x]; Let t = tail[x];
Step 4. /* Test t */
    If t = nil then
        Return[join[succ[h],t]]; /* increment head and enclose in () */
Step 5. /* test t for wrap-around */
    Let u = up[t]; If u = nil then /* no wrap-around */
        Return[join[h,succ[t]]]; /* increment tail and join head back to it */
Step 6. /* Know t wrapped around to u; test h for wrap-around */
    Let v = up[h]; If v = nil then /* no wrap-around */
        Return[join[succ[h],t]]; /* increment head and join it back to tail */
Step 7. /* Both head and tail wrap around. */
    Return[join[v,tail[tail[u]]]]; /* increase size of head by one at cost to tail
*/
```

---

We could also specify an algorithm for a predecessor, returning 0 from argument 0, and from that and the above function, build up arithmetic by induction in the usual way, just to show it can be done, but we won't in this report. Again, all bsxes correspond to numbers, the map $d$ taking $BSX_*$ one-to-one onto $\mathbb{N}$ .

# 10. Summary

We have shown an explicit encoding of the natural numbers as nested parentheses strings, and thence we considered the associated prefix-free binary encoding. We showed that the BSX code is asymptotically optimal. We explored a family of probability distributions on $\mathbb{N}$ with the curious property of their being isomorphic to their random sampling space. We described the programming language BILL, a minimal version of Lisp, and showed how the encoding gives rise to a system of Gödel numbering for partial recursive functions.

## Appendix: The Sampling Space Construction

Let $p$ be an arbitrary probability measure, so that the domain of $p$ is a $\sigma$-algebra $\mathcal{E}$ of "events" whose union is the event $\Omega$ consisting of all "elementary outcomes". Let $(X,\mathcal{B})$ be a measurable space (i.e., $\mathcal{B}$ is a $\sigma$-algebra of subsets of X), and F: $\Omega \overset{\square}{\to} X$ a measurable map, so that $F^{-1}[B]\in\mathcal{E}$ whenever $B\in\mathcal{B}$ . Then $F_*$ maps probability measures on $\Omega$ to probability measures on X:

$F_*(p)(B) = p(F^{-1}[B])$

$F_*(p)$ is the distribution of the X-valued r.v. F.

$\forall\, m > 0$, $p^{\otimes m}$ is the $m$-fold product measure on $\Omega^m$: $p^{\otimes m}(\overset{m-1}{\underset{j=0}{\times}} E_j) = \prod_{j=0}^{m-1} p(E_j)$ . Its domain is the $\sigma$-algebra $\mathcal{E}^{\otimes m}$ of subsets of $\Omega^m$ generated by the "boxes" $\overset{m-1}{\underset{j=0}{\times}} E_j$, where $E_j\in\mathcal{E}$ for all $j\in m$ . $p^{\otimes 0}$ is the Dirac measure concentrated on $\{\Lambda\}$, where $\Lambda \in \Omega^*$ is the empty string.

Let $(\mu_j)$ be a fixed family of probability weights (non-negative numbers summing to 1). Given a probability measure $p$, its sampling measure [weighted according to $\mu$] is the measure $\oplus_m \mu_m p^{\otimes m}$ on the set of lists $\Omega^*$ of elements of $\Omega$. Its domain is the $\sigma$-algebra $\{A\subseteq\Omega | \forall\, m\in\mathbb{N}, A \cap \Omega^m \in \mathcal{E}^{\otimes m}\}$, and its value is given by

$(\oplus_m \mu_m p^{\otimes m})(A) = \sum_{m=0}^{\infty} \mu_m p^{\otimes m}(A \cap \Omega^m)$ . On $\Omega^*$ the discrete r.v. $M(\xi)=\#\xi$ has the distribution determined by $\mu$ :

$(\oplus_m \mu_m p^{\otimes m})(M=n) = \mu_n$ . A point in $\Omega^*$ represents a random sample from the population $p$, where the sample size $M$ is itself a r.v., with distribution determined by $\mu$. The conditional probability $(\oplus_m \mu_m p^{\otimes m})(\;|M=n) = p^{\otimes n}$

## Appendix: Some Programming Considerations

### ■ Efficient Algorithms for Numerical Joining and Separation

It is straightforward to convert the definitions of numerical separation and join into efficient programs provided there are efficient procedures for computing the Catalan numbers and the consecutive sums thereof. The recursive formula (3) was critical to our definition of the bsx encoding and decoding, but more efficient ones exist.

Staring from (5), we have for $n > 0$, $C_n = \frac{\binom{2n}{n}}{n+1} = \frac{(2n)(2n-1)(2(n-1))!}{(n+1)n^2((n-1)!)^2} = (4 - \frac{6}{n+1})C_{n-1}$. This relation is the basis for iteration algorithms such as the following:

---

```
Step 0. /* The input is a number n, and the output is the pair (cn, sn) where cn is
the nth Catalan number and sn is the sum of the Catalan numbers from 0 to n - 1 . k
is a loop index and t holds the next sum. */
    Input n;
Step 1. /* Dispose of special cases */ If n = 0 then output (1, 0) and return; If n
= 1 output (1, 1) and return.
    Step 2. /* n > 2; initialize for loop */  Let cn = 1; Let sn = 2; Let k = 2.
Step 3. /* update */  Let cn = cn*(4 - 6/(k + 1)); Let t = sn + cn .
     Step 4. /* Test and exit  */ If k = n then output (cn, sn) and return.
     Step 5. /* update sum  and loop */
      Let sn = t; Let k = k + 1; Go to Step 3.
```

---

In practice the first 20 Catalan numbers, at least, might be stored in an internal table. Note $C_{19}$ = 1,767,263,190 < $2^{32}$ < $C_{20}$ = 6,564,120,420 .

# References

[AIT] G.J. Chaitin, *Algorithmic Information Theory*, Cambridge University Press, 1990, ISBN 0-521-34306-2. See also http://www.cs.auckland.ac.nz/~chaitin/
for his latest version of LISP.

[ConMath] Graham, Knuth, Patashnik, *Concrete Mathematics*, Addson-Wesley, 1989, ISBN0-201-14236-8

[Dyck] http://en.wikipedia.org/wiki/Dyck_language

[Elias] http://en.wikipedia.org/wiki/Elias_delta_coding

[RDO] http://www.cs.txstate.edu/~ro01/bsxCodeTalk/
is a URL which contains implementations of the algorithms outlined in this report, as *Mathematica* functions.